\documentclass[10pt,twocolumn]{article}

\usepackage[a4paper,margin=2cm,columnsep=0.8cm]{geometry}
\usepackage{amsmath,amssymb}
\usepackage{graphicx}
\usepackage{booktabs}
\usepackage{xcolor}
\usepackage[colorlinks=true,linkcolor=blue!50!black,citecolor=blue!50!black,urlcolor=blue!50!black]{hyperref}

\newcommand{\stdweb}{\textsc{STDWeb}}
\newcommand{\stdpipe}{\textsc{STDPipe}}
\newcommand{\gaia}{\textit{Gaia}}
\newcommand{\mmag}{\,\mathrm{mmag}}
\newcommand{\msys}{m_{\rm sys}}

\title{\bfseries Multi-night photometric repeatability of \stdweb:\\
an empirical error budget from a 14-night campaign on a single field}

\author{P.-Y.~Lechapelain\\[2pt]
\normalsize Observatoire de Bretagne Sud (IAU station Y80), Vannes, France\\
\normalsize \texttt{obs.south.brittany@gmail.com}}

\date{\today}

\begin{document}
\maketitle

% ============================================================
\begin{abstract}
\noindent
Web-based photometric pipelines such as \stdweb{} \cite{Karpov2025} are now
routinely used by small observatories and professional--amateur (pro--am)
networks across the full range of photometric science --- variable stars,
supernovae, exoplanet transits, and the follow-up of high-energy transients.
The formal magnitude uncertainty they report describes a single reduction of a single frame; campaign science
instead requires the scatter of the \emph{same star, same field, night after
night}. We measure this empirically using the field of the Einstein Probe
trigger EP-WXT~01709274151, observed on 14 nights between 2026 July 6 and
July~24: 157 \gaia-like $G$ exposures reduced homogeneously with \stdweb{},
analysed through 101 constant field stars ($10.5<G<14.5$). The median formal
error is $\sigma_{\rm form}=8.1\mmag$ ($6.7\mmag$ for the bright half), and
within a single night the empirical differential scatter matches it
($\chi_{\rm within}\simeq1$). Used naively --- taking the pipeline's
\texttt{mag\_calib}, which lives in the instrumental pseudo-band --- each
star acquires a night-level offset of $\sigma_{\rm night}\simeq7$--$9\mmag$,
the campaign-level single-epoch scatter grows to $11$--$12\mmag$, and the
formal error is roughly \emph{half} of the true one
($\chi_{\rm camp}=1.5$--$1.7$). We show that this entire excess is the
epoch-dependent colour term: analysing the catalogue-system magnitude
$\msys=\texttt{mag\_calib}+\mathrm{colour\ term}\times(BP\!-\!RP)$ collapses
the night term to $3.4\mmag$, restores $\chi_{\rm camp}=1.0$--$1.1$, and
improves the absolute \gaia-anchored per-star repeatability from $43\mmag$ to
$9.5\mmag$, with the nightly zero-point peak-to-peak shrinking from
$130\mmag$ to $11\mmag$ and the airmass and colour correlations of the
residuals vanishing. Stellar colours fitted directly from the light curves
--- as the slope of \texttt{mag\_calib} against the per-epoch colour term ---
reproduce \gaia{} $BP\!-\!RP$ to $0.027$\,mag ($r=0.98$) and yield an
identical budget, so the correction requires no external colour catalogue.
We provide the closed variance decomposition, practical recommendations, and
a cautionary note on re-deriving zero points outside the pipeline.
\end{abstract}

% ============================================================
\section{Introduction}

Standardized, publicly available photometric pipelines have changed how
small observatories produce science-grade photometry, for applications
ranging from variable stars, supernovae and exoplanet transits to the
follow-up of high-energy transients. Their value is methodological as much
as practical: instead of each observer implementing an independent reduction
chain --- a step where independent implementations can silently introduce
systematic errors --- a common pipeline provides a uniform, versioned and
reproducible reduction across heterogeneous instruments, reducing the margin
for human error \cite{Irving2026}. \stdweb{} \cite{Karpov2025}, the web
front-end to the \stdpipe{} library \cite{Karpov2021}, implements such a
workflow --- masking, source extraction, astrometric calibration, photometric
solution against survey catalogues with optional colour term and positionally
varying zero point, image subtraction and forced photometry --- behind a
browser interface requiring no local software. The attention paid to
pipeline-level systematics in this community is illustrated by the GRANDMA
network, which has reduced common campaigns in parallel with \stdpipe{} and
the independent Muphoten pipeline \cite{Duverne2022} precisely to verify
that the two chains produce consistent results \cite{Aivazyan2022}.

For a single image, \stdweb{} reports a calibrated magnitude and a formal
uncertainty (\texttt{mag\_calib\_err}) combining the flux measurement error
with the local zero-point model error. This answers ``how well was this star
measured on this frame?''. Campaign science asks a different question: ``if I
measure the same star on the same field night after night, what scatter should
I expect?''. The two differ by every effect that is coherent within one frame
or one night but varies between nights: transparency and airmass changes,
colour-dependent extinction, zero-point model residuals, seeing- and
aperture-dependent flux losses.

That empirical scatter in ground-based time-series photometry exceeds the
formal, white-noise expectation is well documented: correlated systematics
(``red noise'') degrade detection thresholds and parameter uncertainties
well beyond photon statistics \cite{Pont2006}, and dedicated detrending
algorithms such as SysRem \cite{Tamuz2005} and TFA \cite{Kovacs2005} exist
precisely to remove ensemble-level systematics that formal errors do not
describe. More recently, work on automated pipelines has stressed that
photometric calibration accuracy and the calibration of the reported
uncertainties are distinct questions that must be validated separately
\cite{Erece2026}. For \stdweb{} specifically, published validation has
focused on single-epoch calibration and on cross-pipeline consistency over
common data sets \cite{Aivazyan2022}; to our knowledge, no dedicated
multi-night repeatability budget --- same field, same instrument, many
nights --- has been published. The EP-WXT~01709274151 follow-up campaign at
the Observatoire de Bretagne Sud provides exactly such a data set: one
field, one instrument, one pipeline, 14 nights; the transient follow-up
context matters only in that it set the observing cadence. We stress that the science target itself plays no role
here: this paper uses only the constant field stars, as a measurement of the
\emph{system} --- telescope, camera, atmosphere and pipeline together.

Section~\ref{sec:obs} describes the observations; Section~\ref{sec:method}
the ensemble and estimators; Section~\ref{sec:results} the measured budget,
including the decisive role of the colour term;
Section~\ref{sec:discussion} the interpretation;
Section~\ref{sec:conclusions} practical recommendations.

% ============================================================
\section{Observations and data reduction}
\label{sec:obs}

\subsection{Instrument and site}
Observations were obtained at the Observatoire de Bretagne Sud (IAU station
Y80, southern Brittany, France) with a 254\,mm $f/3$ Newtonian reflector
and an SVBONY SV605MC camera (Sony IMX533 back-illuminated colour CMOS),
operated at $-4.8^{\circ}$C, gain 5. The measured pixel scale is
$1.05''$\,px$^{-1}$ for a $52'\times52'$ field of view; the median stellar
FWHM over the campaign is $\simeq4''$ ($3.9$\,px). The delivered image
sampling and the colour sensor make this a representative low-cost pro--am
setup.

\subsection{The field}
The field is that of the Einstein Probe Wide-field X-ray Telescope trigger
01709274151, whose optical counterpart --- the flaring high-proper-motion
M~dwarf PM~J15587+2351E --- was announced in GCN Circular 45133. The target
and its common-proper-motion companion (the only two ensemble candidates with
$\mu>100$\,mas\,yr$^{-1}$) are explicitly \emph{excluded} from the comparison
ensemble; everything that follows concerns the constant field stars only.

\subsection{Campaign and reductions}
\label{sec:reductions}
The analysed data set comprises 157 $G$-band exposures on 14 nights between
2026 July~6 and July~24 (the campaign also collected 153 $BP$ and 148 $RP$
exposures, not used here). The exposure mix is 11 frames of 120\,s (first
night) and 146 frames of 30\,s (monitoring phase); airmass spans 1.11--2.78.
Six epochs yield no unflagged ensemble detections and drop out of the
statistics, leaving 151 usable epochs.

All frames were reduced homogeneously with a self-hosted \stdweb{} instance:
SExtractor \cite{Bertin1996} source extraction via \stdpipe, astrometric
calibration, and a photometric solution against \gaia{} \cite{GaiaDR3} $G$
with a $BP\!-\!RP$ colour term and a positionally varying zero point. For
each frame the pipeline delivers a source table with calibrated magnitudes
(\texttt{mag\_calib}), formal errors (\texttt{mag\_calib\_err}) and the
fitted colour-term coefficient $c_t$ of that frame.

One point deserves emphasis, because it drives the central result of this
paper: \texttt{mag\_calib} is expressed in the \emph{instrumental
pseudo-band} of the system. The magnitude in the catalogue system is
\begin{equation}
\msys \;=\; \texttt{mag\_calib} \;+\; c_t\,(BP\!-\!RP),
\label{eq:msys}
\end{equation}
where $c_t$ is the per-epoch colour-term coefficient and $BP\!-\!RP$ the true
colour of the star \cite{KarpovPC}. Over this campaign $c_t$ has median
$-0.194$ with a night-to-night scatter of $0.053$ and nightly medians ranging
from $-0.14$ to $-0.28$ (Fig.~\ref{fig:colorterm}, left): the transformation
between the instrumental and catalogue systems is \emph{not constant in
time}. All statistics below are therefore computed twice --- on
\texttt{mag\_calib} (the naive use) and on $\msys$ --- directly from the
per-frame tables; no zero point is re-derived outside the pipeline
(Sect.~\ref{sec:caution}).

% ============================================================
\section{Comparison ensemble and estimators}
\label{sec:method}

\subsection{Ensemble selection}
Sources of every frame were matched to the \gaia{} reference catalogue of
their task within $1.5''$, after propagating \gaia{} proper motions to the
epoch of observation. The ensemble retains stars with (i) SExtractor
\texttt{flags}=0, (ii) $10.5<G<14.5$ (photon noise subdominant, no
saturation), (iii) detection on at least 25\% of epochs, and (iv)
$\mu<100$\,mas\,yr$^{-1}$ (removing the science target and its companion).
101 stars pass all cuts. The ensemble spans $\sim$4\,mag, so all stability
statistics are per star; the ``bright half'' below denotes the 51 stars
brighter than the ensemble median.

\subsection{Estimators}
Let $m_{i,k}$ be the magnitude of star $i$ on epoch $k$ (either
\texttt{mag\_calib} or $\msys$), belonging to night $n(k)$, and $G_i$ its
\gaia{} $G$ magnitude.

\paragraph{Formal error.} $\sigma_{\rm form}$: ensemble median of
\texttt{mag\_calib\_err}.

\paragraph{Differential residual.} With $\bar m_i$ the campaign median of
star $i$ and the per-epoch ensemble offset
$c_k=\operatorname*{median}_j\,(m_{j,k}-\bar m_j)$,
\begin{equation}
\delta_{i,k}=(m_{i,k}-\bar m_i)-c_k .
\end{equation}
This is classical ensemble differential photometry: any common-mode term of
epoch $k$ cancels exactly.

\paragraph{Within-night scatter.}
$\sigma_{\rm within}$: per star, the median over nights of
$\operatorname{std}_k(\delta_{i,k})$ computed inside each night
($\geq5$ epochs); ensemble median thereof.

\paragraph{Night term.}
$\sigma_{\rm night}$: per star, the standard deviation over nights of the
nightly mean $\langle\delta_{i,k}\rangle_{k\in n}$ --- the star-level offset
that survives ensemble correction from one night to the next.

\paragraph{Campaign scatter.}
$\sigma_{\rm camp}$: per star, $\operatorname{std}_k(\delta_{i,k})$ pooled
over the whole campaign. To a good approximation
$\sigma_{\rm camp}^2\simeq\sigma_{\rm within}^2+\sigma_{\rm night}^2$.

\paragraph{\gaia-anchored residual.} For absolute stability: nightly zero
point $Z_n=\operatorname*{median}_{i,k\in n}(m_{i,k}-G_i)$ and residual
$r_{i,k}=m_{i,k}-G_i-Z_{n(k)}$, with per-star RMS $\sigma_i$.

\subsection{Data-driven stellar colours}
\label{sec:fitcolor}
\begin{figure*}[t]
\centering
\includegraphics[width=0.9\textwidth]{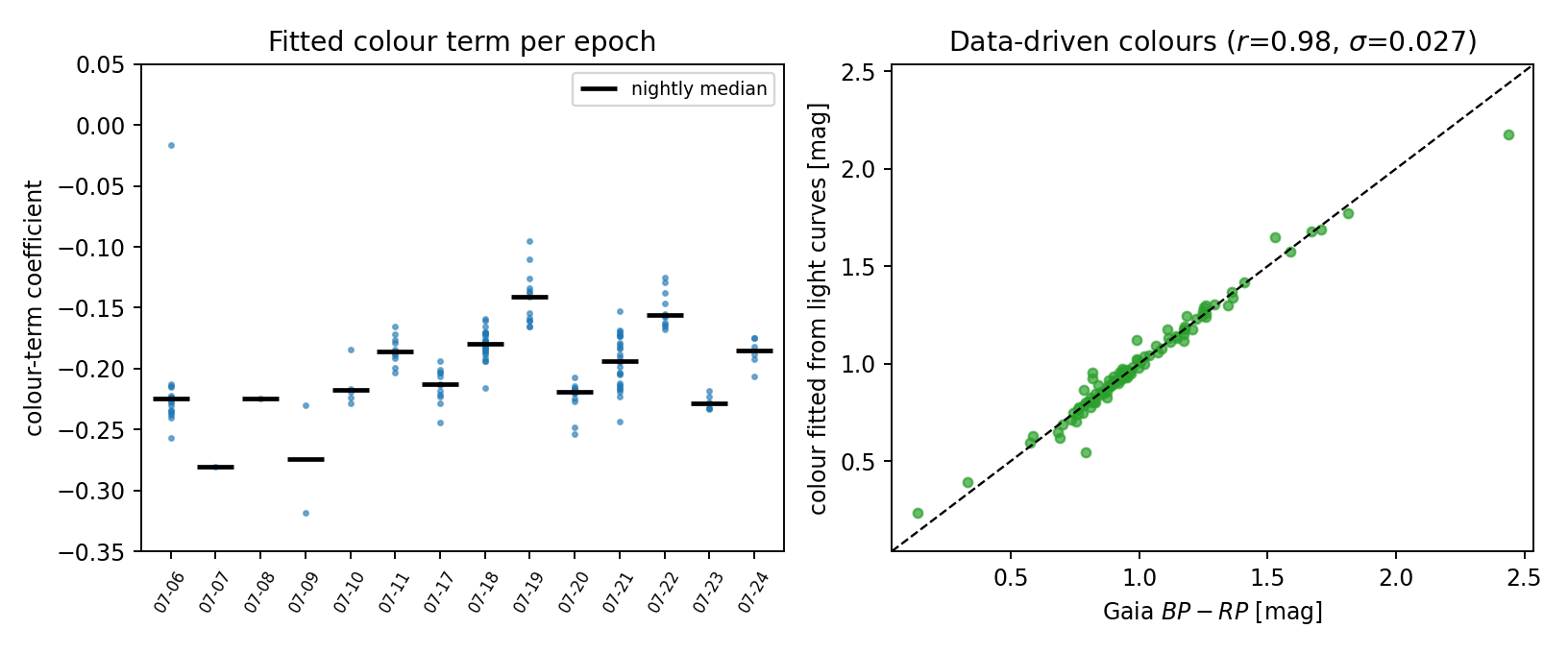}
\caption{\emph{Left:} fitted colour-term coefficient $c_t$ per epoch over the
campaign; the nightly medians move between $-0.14$ and $-0.28$, so the
instrumental-to-catalogue transformation is time-dependent. \emph{Right:}
stellar colours fitted from the light curves as the slope of
\texttt{mag\_calib} versus $c_t$, against \gaia{} $BP\!-\!RP$; the identity
line is dashed.}
\label{fig:colorterm}
\end{figure*}

Equation~(\ref{eq:msys}) needs the true colour of each star. When no
catalogue colour is available, it can be fitted from the data themselves: the
colour term is set by the observing conditions and is physically uncorrelated
with stellar variability, so the colour of a constant star is simply
(minus) the slope of its \texttt{mag\_calib} light curve against the
per-epoch coefficient $c_t$ \cite{KarpovPC}. We implement this as an
iteratively $3\sigma$-clipped linear regression of $m_{i,k}$ on $c_{t,k}$
per star. Over the 101 ensemble stars the fitted colours reproduce \gaia{}
$BP\!-\!RP$ with $r=0.98$, zero median offset ($0.002$\,mag) and a robust
scatter of $0.027$\,mag (Fig.~\ref{fig:colorterm}, right) --- from $G$-band
photometry alone.

% ============================================================
\section{Results}
\label{sec:results}

\begin{table}[t]
\centering
\caption{Error budget on the naive \texttt{mag\_calib} (ensemble medians;
brackets: 16--84\% range over stars). Values in mmag.}
\label{tab:budget}
\setlength{\tabcolsep}{4pt}
\resizebox{\linewidth}{!}{%
\begin{tabular}{@{}lcc@{}}
\toprule
Quantity & All (101) & Bright half (51) \\
\midrule
$\sigma_{\rm form}$ & 8.1 & 6.7 \\
$\sigma_{\rm within}$ & 8.1 [6.3--11.5] & 6.8 [5.8--8.7] \\
$\sigma_{\rm night}$ & 6.7 [3.8--15.7] & 8.6 [4.5--19.4] \\
$\sigma_{\rm camp}$ & 11.8 [8.2--22.1] & 11.3 [7.6--22.1] \\
$\chi_{\rm within}=\sigma_{\rm within}/\sigma_{\rm form}$ & 1.0 & 1.0 \\
$\chi_{\rm camp}=\sigma_{\rm camp}/\sigma_{\rm form}$ & 1.46 & 1.69 \\
\midrule
$\sigma_{\rm within}$, 120\,s frames & \multicolumn{2}{c}{4.1 [2.5--5.8]} \\
$\sigma_{\rm within}$, 30\,s frames & \multicolumn{2}{c}{7.7 [6.1--10.9]} \\
\bottomrule
\end{tabular}}
\end{table}

\begin{table}[t]
\centering
\caption{Effect of the colour term: same estimators on the naive
\texttt{mag\_calib} versus the catalogue-system magnitude $\msys$
(Eq.~\ref{eq:msys}) with \gaia{} colours, and with colours fitted from the
data (Sect.~\ref{sec:fitcolor}). Ensemble medians, mmag.}
\label{tab:msys}
\setlength{\tabcolsep}{4pt}
\resizebox{\linewidth}{!}{%
\begin{tabular}{@{}lccc@{}}
\toprule
Quantity & \texttt{mag\_calib} & $\msys$ (\gaia) & $\msys$ (fitted) \\
\midrule
$\sigma_{\rm within}$ (bright) & 6.8 & 6.2 & 6.2 \\
$\sigma_{\rm night}$ (bright) & 8.6 & 3.4 & 3.2 \\
$\sigma_{\rm camp}$ (bright) & 11.3 & 6.9 & 6.9 \\
$\chi_{\rm camp}$ (bright) & 1.69 & 1.03 & 1.03 \\
$\sigma_{\rm camp}$ (all) & 11.8 & 8.6 & 8.7 \\
\midrule
\gaia-anchored per-star RMS & 43 [32--56] & 9.5 [7.0--12.0] & 9.4 \\
Nightly ZP $\sigma(Z_n)$ / ptp & 38 / 130 & 2.3 / 11 & 2.4 / 10 \\
$\operatorname{corr}(Z_n,\,\overline{\rm airmass}_n)$ & $-0.79$ & $-0.18$ & --- \\
$\operatorname{corr}(\bar r_i,\,BP\!-\!RP)$ & $+0.60$ & $+0.05$ & --- \\
\bottomrule
\end{tabular}}
\end{table}

\begin{figure}[t]
\centering
\includegraphics[width=\linewidth]{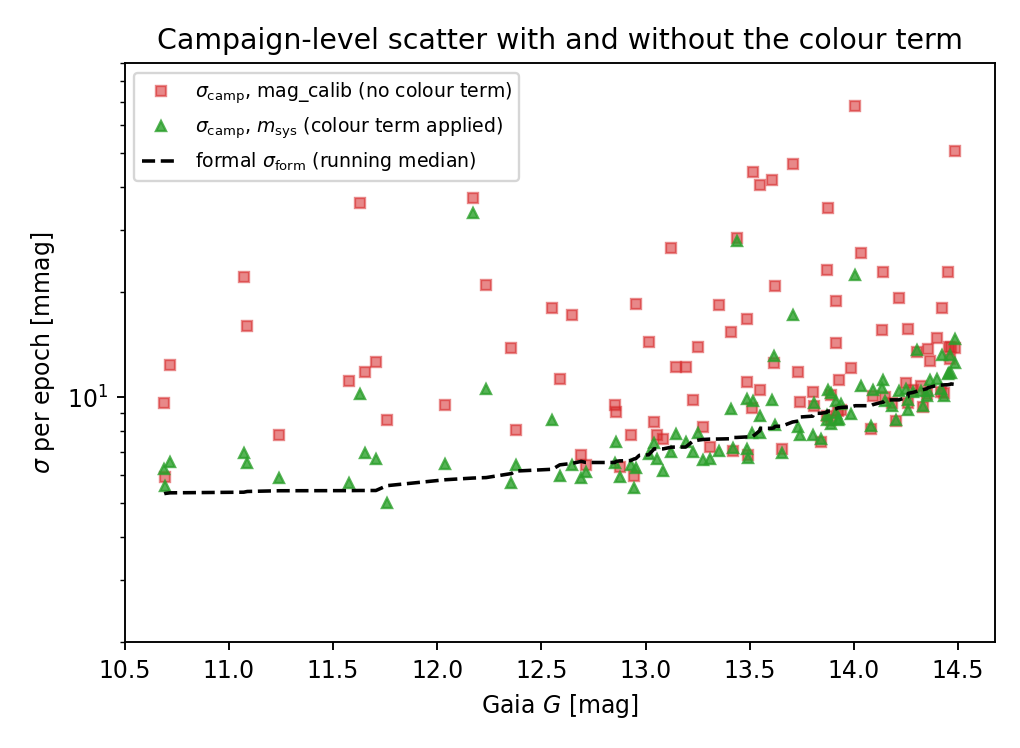}
\caption{Per-star campaign-level single-epoch scatter versus \gaia{} $G$, on
the naive \texttt{mag\_calib} (squares) and on the catalogue-system
magnitude $\msys$ (triangles), with the running median of the formal error
(dashed). Applying the colour term collapses the campaign scatter of the
bright stars onto the formal-error line.}
\label{fig:scatter}
\end{figure}

\begin{figure*}[t]
\centering
\includegraphics[width=0.96\textwidth]{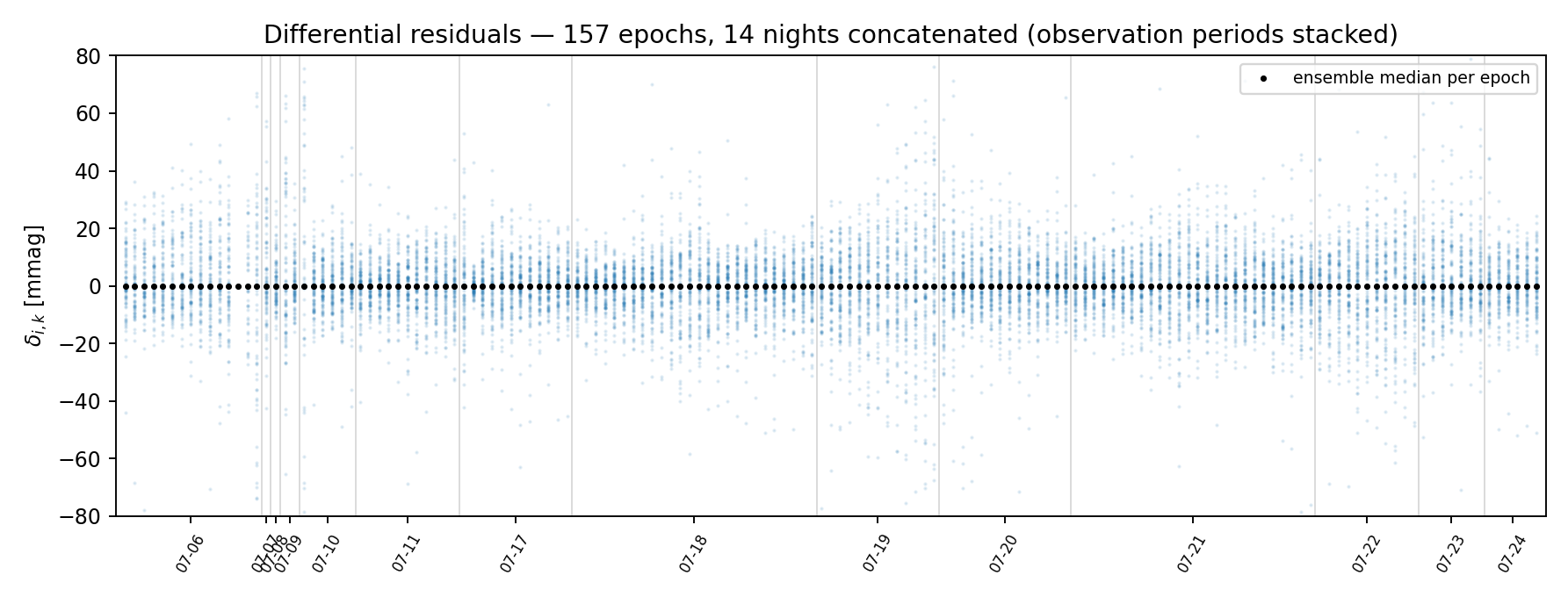}
\caption{Differential residuals $\delta_{i,k}$ of the 101 ensemble stars (on
\texttt{mag\_calib}) with the 14 nights concatenated (inter-night gaps
suppressed; vertical lines mark night boundaries). Black points: per-epoch
ensemble median. The band stays flat at the $\sim$10\,mmag level; the small
star-level night offsets visible here constitute $\sigma_{\rm night}$ and
are removed by the colour term (Table~\ref{tab:msys}).}
\label{fig:stacked}
\end{figure*}

\subsection{Formal errors and within-night scatter}
The median formal error is $\sigma_{\rm form}=8.1\mmag$ over the full
ensemble and $6.7\mmag$ for the bright half. The empirical within-night
differential scatter is statistically identical: $8.1\mmag$ and $6.8\mmag$
respectively, i.e.\ $\chi_{\rm within}\simeq1.0$
(Table~\ref{tab:budget}). The exposure-time strata behave as photon
statistics predict: $4.1\mmag$ on the 120\,s frames versus $7.7\mmag$ on the
30\,s frames, a ratio of 1.9 for a $4\times$ exposure ratio. \emph{Within a
single night, \stdweb{} formal errors are an accurate description of
differential precision.}

\subsection{The naive multi-night budget}
Used naively --- pooling \texttt{mag\_calib} across nights --- each star
acquires its own offset: the nightly-mean residual moves by
$\sigma_{\rm night}=6.7\mmag$ (median; $8.6\mmag$ for the bright half, where
photon noise no longer hides it) from night to night, even though the
ensemble correction removes the common mode exactly. Added in quadrature
with the within-night term this reproduces the directly measured campaign
scatter ($\sqrt{6.8^2+8.6^2}=11.0$ vs the measured $11.3\mmag$ for the
bright half): the budget closes. The campaign-level single-epoch
uncertainty is then $\sigma_{\rm camp}=11.8\mmag$ (all) and $11.3\mmag$
(bright), giving $\chi_{\rm camp}=1.46$ and $1.69$ respectively: on
\texttt{mag\_calib}, the formal error is roughly half of the real one.

\subsection{The colour term is the night term}
\label{sec:colorterm}
The fitted colour-term coefficient varies from night to night
(Fig.~\ref{fig:colorterm}, left): $\sigma(c_t)=0.053$ around a median of
$-0.194$. A star whose colour differs from the ensemble reference by
$\Delta(BP\!-\!RP)\sim0.3$\,mag therefore picks up night-level excursions of
order $0.05\times0.3\simeq15\mmag$ in \texttt{mag\_calib} --- precisely the
scale and the colour signature of the measured night term.

Analysing the catalogue-system magnitude $\msys$ (Eq.~\ref{eq:msys})
confirms the diagnosis quantitatively (Table~\ref{tab:msys},
Fig.~\ref{fig:scatter}): the night term collapses from $8.6$ to $3.4\mmag$
(bright half), the campaign scatter from $11.3$ to $6.9\mmag$, and
\begin{equation}
\chi_{\rm camp}(\msys)=1.03\ \text{(bright)},\qquad 1.07\ \text{(all)} .
\end{equation}
The absolute \gaia-anchored per-star repeatability improves from $43$ to
$9.5\mmag$, the nightly zero point steadies from
$\sigma(Z_n)=38\mmag$ (peak-to-peak $130\mmag$) to $2.3\mmag$ (peak-to-peak
$11\mmag$), and the two systematic correlations that betrayed the residual
chromatic term --- $Z_n$ with airmass ($-0.79$) and star residuals with
$BP\!-\!RP$ ($+0.60$) --- essentially vanish ($-0.18$ and $+0.05$).

Repeating the entire budget with colours fitted from the data
(Sect.~\ref{sec:fitcolor}) instead of \gaia{} colours gives an
indistinguishable result (last column of Table~\ref{tab:msys}): the
correction is self-contained and does not require an external colour
catalogue.

% ============================================================
\section{Discussion}
\label{sec:discussion}

\subsection{A closed, and now explained, budget}
The campaign supports the decomposition
\begin{equation}
\sigma^2_{\rm camp}=\underbrace{\sigma^2_{\rm within}}_{\simeq\;\sigma^2_{\rm form}}
+\;\sigma^2_{\rm night},
\end{equation}
and identifies $\sigma_{\rm night}$ on \texttt{mag\_calib} as, dominantly,
the epoch-dependent colour term acting on each star's colour offset from the
ensemble. The formal error is not wrong --- it fully accounts for the
intra-night noise --- and once magnitudes are expressed in the catalogue
system the budget is complete: $\chi_{\rm camp}(\msys)\simeq1$. The residual
night term on $\msys$ ($\simeq3\mmag$) plausibly contains second-order
chromatic effects, flat-field/illumination residuals and low-level aperture
systematics, and sets the current floor of the system at
$\sim$7\,mmag per epoch over a campaign.

\subsection{Formal errors are half the real ones --- until the colour term
is applied}
The practical headline is two-sided. On the naive \texttt{mag\_calib},
$\chi_{\rm camp}=1.5$--$1.7$: a multi-night light curve carries a true
single-epoch uncertainty roughly \emph{twice} the reported formal error for
bright ensemble stars, and any $\chi^2$-based variability statistic,
period-search false-alarm probability, or flare-amplitude error bar computed
from raw \texttt{mag\_calib\_err} is overconfident by a factor 2.5--3 in
variance. But the excess is not an irreducible property of the system: it is
the missing half of the photometric solution. Applying
Eq.~(\ref{eq:msys}) --- with catalogue colours or with colours fitted from
the data --- restores $\chi_{\rm camp}\simeq1$, and formal errors can then be
used as-is even across nights. Users who cannot apply the colour term (e.g.\
targets without colour information and too few epochs to fit one) should
inflate formal errors by $\chi_{\rm camp}\simeq1.5$--$2$ or add an
$8\mmag$ night term in quadrature.

\subsection{Absolute photometry at the 10\,mmag level}
On $\msys$, absolute magnitudes anchored to \gaia{} $G$ repeat at
$9.5\mmag$ per star across 14 nights spanning airmass 1.1--2.8, with a
nightly zero point stable to $2.3\mmag$ --- an order of magnitude better
than the naive $43\mmag$ floor, and achieved by a low-cost colour-CMOS
system with no photometric-night selection. For a pro--am station this is
the difference between contributing relative light curves and contributing
calibrated photometry directly comparable across the network.

\subsection{A cautionary tale: do not re-derive zero points}
\label{sec:caution}
During this work, an initial reduction that reconstructed zero points outside
the pipeline (instead of using the pipeline's calibrated magnitudes directly)
produced apparent nightly zero-point excursions of up to 0.7\,mag and a
spurious per-star repeatability of 0.15--0.2\,mag --- numbers that would have
condemned the system unfairly, and that disappeared entirely once the
analysis was rerun on the pipeline's own outputs. The failure mode (an
exposure-time-dependent zero-point error applied to the 30\,s frames) was
only caught by spot-checking individual stars against the per-frame source
tables. We recommend that any campaign-level meta-analysis of \stdweb{}
products treat \texttt{mag\_calib} and $c_t$ as the primary observables and
validate any re-derived quantity against them star by star.

\subsection{Caveats}
(1) One field, one instrument, one summer: the night term and its chromatic
origin should be re-measured per station; systems with photometric filters
will show a smaller $c_t$ variance. (2) \gaia{} $G$ as reference: a small
part of the absolute floor may remain bandpass mismatch rather than
instability. (3) The exposure mix (11$\times$120\,s, 146$\times$30\,s) is
unbalanced; the 120\,s stratum comes from a single night. (4) Ensemble
members are assumed constant; low-level intrinsic variability of individual
stars inflates the upper percentiles of $\sigma_{\rm camp}$ but not its
median.

% ============================================================
\section{Conclusions}
\label{sec:conclusions}

From 157 exposures of one field on 14 nights, using 101 constant field stars
and treating the telescope--camera--atmosphere--pipeline chain as a single
system:

\begin{enumerate}
\item \stdweb{} formal errors are accurate \emph{within a night}:
$\sigma_{\rm within}\simeq\sigma_{\rm form}=7$--$8\mmag$,
$\chi_{\rm within}\simeq1$.
\item Pooling the pipeline's \texttt{mag\_calib} naively across nights, a
star-level night term of $7$--$9\mmag$ appears and the campaign-level
single-epoch uncertainty grows to $11$--$12\mmag$: \emph{the formal error
is then roughly half of the real one} ($\chi_{\rm camp}=1.5$--$1.7$).
\item That night term is the epoch-dependent colour term. Analysing the
catalogue-system magnitude
$\msys=\texttt{mag\_calib}+c_t\,(BP\!-\!RP)$ collapses it to $3.4\mmag$ and
restores $\chi_{\rm camp}=1.0$--$1.1$; formal errors are then valid across
nights.
\item On $\msys$, absolute \gaia-anchored photometry repeats at
$9.5\mmag$ per star over 14 nights (nightly zero point stable to
$2.3\mmag$), versus $43\mmag$ naively.
\item Stellar colours fitted from the light curves themselves (slope of
\texttt{mag\_calib} vs $c_t$) match \gaia{} $BP\!-\!RP$ to $0.027$\,mag and
give an identical budget: the correction needs no external colour catalogue.
\item Meta-analyses of \stdweb{} products should consume
\texttt{mag\_calib} and $c_t$ directly; re-deriving zero points is the
dominant avoidable failure mode.
\end{enumerate}

These numbers give pro--am networks a concrete, closed and fully explained
error model for \stdweb-based follow-up, and an estimator set that any
station can run on its own task archive.

% ============================================================
\section*{Acknowledgements}
This work uses \stdweb{} and \stdpipe{} by Sergey Karpov, and data products
of the \gaia{} mission. We warmly thank Sergey Karpov for pointing out that
the catalogue-system magnitude, not \texttt{mag\_calib}, is the quantity
whose stability should be analysed, and for suggesting the data-driven
colour-fitting method of Sect.~\ref{sec:fitcolor}. The campaign was
conducted within the RAPAS pro--am network in response to Einstein Probe
alerts. Data analysis and manuscript preparation were assisted by Claude
(Anthropic); all analysis choices, verifications and conclusions are the
author's responsibility.

% ============================================================

\end{document}